\documentstyle[preprint,aps]{revtex}

\begin{document}
\draft
\title{A New Approximation Scheme in Quantum
       Mechanics}

\author{Sang Koo You, Kwang Joe Jeon, and Chul Koo Kim\footnote{e-mail:
        ckkim@phya.yonsei.ac.kr}}
\address{Department of Physics and Institute for Mathematical Science,
         Yonsei University, Seoul 120-749, Korea}
\author{Kyun Nahm}
\address{Department of Physics, Yonsei University, Wonju 220-710,
         Korea}
\maketitle

\begin{abstract}
An approximation method which combines the perturbation theory with
the variational calculation is constructed for quantum mechanical
problems. Using the anharmonic oscillator and the He atom as examples,
we show that the present method provides an efficient scheme in
estimating both the ground and the excited states. We also discuss
the limitations of the present method.
\end{abstract}


\section{Introduction}
In undergraduate quantum mechanics, students are taught generally
two types of approximation schemes, the perturbation theory and
the variational calculation. Each method is known to have its own advantages and
limitations. The perturbation theory is applicable only when the
perturbing potential is small, whereas the variational method lacks
systematic means for improvement after the first approximation.
Also, it is rather cumbersome and difficult to estimate the excited
states in the variational method \cite{one,two,three,four}.

In this paper, we present an approximate scheme which combines the two
approximation methods. It will be shown that the combined method carries
the advantages of both methods and, at the same time, overcomes the
deficiencies mentioned above. Actually, the concept of combining the
two approximation methods is not entirely new.
For instances, this concept has been used in lattice dynamics to
obtain a better description of phonons in crystals \cite{five}.
In this approach, perturbation calculations were made on the basis
of the variational self-consistent harmonic approximations.
Also, in relativistic field
theories, this approach was called the perturbation with variational
basis and used to calculate effective Gaussian potentials and high
order terms \cite{six,seven}. However, to the authors' knowledge,
this concept has never been discussed in a plain quantum mechanical
language understandable to undergraduate or graduate students who do not have
knowledge of either lattice dynamics or relativistic quantum field theory.
Therefore, we believe that developing this combined approximation
scheme for quantum mechanical problems is not only pedagogically
important, but also useful for real applications in quantum
mechanical problems.

\section{Perturbation with a Variational Basis}
In a variational calculation, we first choose a trial function
$\Psi(\lambda)$ as a function of a variational parameter $\lambda$
for a given Hamiltonian $H$. Then, the ground state energy is estimated
by minimizing the expection value, $\langle \Psi_{\lambda} | H |
\Psi_{\lambda} \rangle$ against $\lambda$. In the present formalism,
we also first choose a trial function $\Psi_{n}(\lambda)$. However,
here, the choice of $\Psi_{n}(\lambda)$ is limited to the cases where
a parent Hamiltonian $H_\lambda$, which carries $\Psi_{n}(\lambda)$
as exact eigenfunctions, can be found. We will discuss implications
of this limitation of the theory below. We just assume here that
such a parent Hamiltonian can be found for a given problem.
Then we rewrite the original Hamiltonian, $H$ as follows,
\begin{eqnarray}
 H &=& H_{\lambda} + H - H_{\lambda} \nonumber \\
   &=& H_{\lambda} + H' \ \ ,
\end{eqnarray}
where $H'= H - H_{\lambda}$ is the perturbing Hamiltonian. Clearly, success of a
perturbative calculation depends on how small $H'$ can be made.
In order to obtain an optimum $H_{\lambda}$, we first determine
$\lambda$ through the condition
\begin{eqnarray}
  \delta\langle\Psi_{n}(\lambda) | H | \Psi_{n}(\lambda) \rangle = 0 .
\end{eqnarray}
Here, we note that this condition is not limited to the ground state,
but is valid to all excited states also. Therefore, generally
$\lambda$ will be dependent on the state number $n$. Thus we express
$\lambda$ and $H_{\lambda}$ as $\lambda_n$ and $H_{\lambda_n}$
respectively. Using the above notations, we carry out the standard
perturbation calculation of $E_n$.
\begin{eqnarray}
  E_n =& & \langle\Psi_{n}(\lambda_{n}) | H_{\lambda_n} |
          \Psi_{n}(\lambda_{n}) \rangle \nonumber \\
       &+&\langle\Psi_{n}(\lambda_{n}) | H' |
          \Psi_{n}(\lambda_{n}) \rangle \nonumber \\
       &+& \sum_{k \neq n} \frac{| \langle\Psi_{k}(\lambda_{n}) | H' |
          \Psi_{n}(\lambda_{n}) \rangle |^{2}}{E_{n}^{(0)}(\lambda_{n})
          -E_{k}^{(0)}(\lambda_{n})}  \nonumber \\
       &+& \cdot \cdot \cdot \ \ .
\end{eqnarray}
Here $E_{n}^{(0)}(\lambda_{n})$ is the $n$th energy eigenvalue of
$H_{\lambda_n}$ ;
\begin{eqnarray}
  H_{\lambda_n} | \Psi_{n}(\lambda_{n}) \rangle =
  E_{n}^{(0)}(\lambda_{n}) | \Psi_{n}(\lambda_{n}) \rangle \ \ .
\end{eqnarray}
Now we compare the above result with the conventional approximation
schemes. First of all, for $n=0$, the first two terms become
\begin{eqnarray}
  \langle\Psi_{0}(\lambda_{0}) | H_{\lambda_0} + H' |
          \Psi_{0}(\lambda_{0}) \rangle
  =\langle\Psi_{0}(\lambda_{0}) | H |
          \Psi_{0}(\lambda_{0}) \rangle \ \ ,
\end{eqnarray}
which is just the conventional variational result of the ground
state energy. The third term is the second order perturbation term
which provides a systematic improvement over the simple variational
ground state energy. We note that the perturbative correction term
is negative for the ground state.
Thus, it pushes down the variational values of the
ground state energy for a better agreement with the true ground state
energy.
However, it should be noted that this correction  might
overcompensate the true ground state energy as observed in the
anharmonic oscillator example in the next chapter. The approximate
energy which combines the variational calculation and the second
order perturbation correction no longer provides an upper bound
to the exact ground state energy.
The perturbation expansion of Eq.(3) is different from the
conventional perturbation theory. The basis functions used in the
perturbation are those obtained through a variational process and the
eigenvalues are also optimized values.

Another important aspect of the present theory is that it can be
readily applied in evaluating excited states unlike the conventional
variational methods. In the usual variational calculation, for the
calculation of an excited state energy, it is necessary to construct a
trial wavefunction which is orthogonal to the ground state
wavefunction, thus making the procedure rather cumbersome and
difficult \cite{four}.
In the present formalism, it may appear that this requirement is
automatically satisfied because $ \Psi_{n}(\lambda_{n}) $ is the
$n$th eigenfunction of $H_{\lambda}$, which is orthogonal to
$ \Psi_{0}(\lambda_{n})$.
However, this point requires a closer examination.
Since $\lambda_n$ is determined through
minimization as given by Eq.(2), it is different for each $n$.
For example, $\lambda_0$ should be used throughout in Eq.(3) for
evaluation of the ground state energy, whereas $\lambda_1$ obtained
from Eq.(2) should be used for
evaluation of the first excited state.
Therefore, the requirement of the orthogonality can be satisfied,
but, in an approximate sense, because $ \Psi_{1}(\lambda_{1})$ is
neither exactly orthogonal to the true ground-state wavefunction
nor to $ \Psi_{0}(\lambda_{0})$.
This point will be made more
clear in the next chapters through examples.

\section{Anharmonic Oscillator}
In this chapter, we apply the above formalism to the anharmonic
oscillator problem which has a $x^4$ term. The Hamiltonian is given by
\begin{eqnarray}
  H = \frac{p^2}{2m} + \frac{m\omega^2}{2}x^{2} + bx^4 \ \ ,
\end{eqnarray}
where $b$ is positive. The conventional perturbative calculation is
possible only when $b$ is small \cite{eight}.
In order to apply the present formalism to this problem, we should
first find a suitable trial function and the corresponding parent
Hamiltonian. A convenient choice is given by
\begin{eqnarray}
  \Psi_{\Omega}(\lambda) =( \frac{m \Omega}{\pi \hbar} )^{\frac{1}{4}}
      e^{-\frac{m \Omega}{2 \hbar}x^2} \ \ ,
\end{eqnarray}
for the ground state trial wavefunction. Here $\Omega$ is a variational
parameter. The corresponding parent Hamiltonian, which carries the
above trial wavefunction as the exact ground state eigenfunction is given by
\begin{eqnarray}
  H_{\Omega} = \frac{p^2}{2m} + \frac{m \Omega^2}{2}x^2 \ \ .
\end{eqnarray}
Using this parent Hamiltonian, we rewrite the original Hamiltonian,
\begin{eqnarray}
  H &=& H_{\Omega} + H' \nonumber \\
    &=& \frac{p^2}{2m} + \frac{m \Omega^2}{2}x^2
       + \frac{m}{2}(\omega^2 - \Omega^2 )x^2 + b x^4 \ \ .
\end{eqnarray}
We denote the $n$th eigenstate of $H_{\Omega}$ by $| n_\Omega \rangle$.
Then the expectation value of
$\langle n_\Omega | H | n_\Omega \rangle$ is given by
\begin{eqnarray}
  \langle n_\Omega | \ &H& | n_\Omega \rangle \nonumber \\
  &=& \langle n_\Omega | H_\Omega | n_\Omega \rangle +
   \langle n | H' | n \rangle  \nonumber \\
    &=& \langle n_\Omega | H_\Omega | n_\Omega \rangle -
    \frac{m(\Omega^2 -\omega^2 )}{2}
    \langle n_\Omega | x^2 | n_\Omega \rangle +
    b\langle n_\Omega | x^4 | n_\Omega \rangle \nonumber \\
   &=& \hbar \Omega (n+ \frac{1}{2}) - \frac{\hbar(\Omega^2 -\omega^2 )}
    {4 \Omega}(2n+1) + \frac{3b \hbar^2}{4m^2 \Omega^2}
    (2n^2 + 2n +1) \ \ .
\end{eqnarray}
Here, we used the standard quantum mechanical results \cite{eight};
\begin{eqnarray}
   \langle n_\Omega | x^2 | n_\Omega \rangle &=&
   \frac{\hbar}{2m \Omega}(2n+1) \ , \nonumber \\
   \langle n_\Omega | x^4 | n_\Omega \rangle &=&
   (\frac{\hbar}{2m \Omega})^2 (6n^2 + 6n +3) \ \ .
\end{eqnarray}
By taking a variation on $\langle n_\Omega | H | n_\Omega \rangle $,
we obtain a relation which determines $\Omega_n$,
\begin{eqnarray}
  \Omega_{n}^{3} - \omega^2 \Omega_n - \frac{6b \hbar}{m^2}
  \frac{2n^2 +2n +1}{2n + 1} = 0 \ \ .
\end{eqnarray}
This equation is valid for any $n$.
Now it is necessary to evaluate the third term of Eq.(3).
This second order calculation
can also be carried out in a straightforward fashion
\cite{eight} to yield
\begin{eqnarray}
  &\sum_{k \neq n}& \frac{|\langle k_\Omega | H' | n_\Omega \rangle|^2}
  {E_{n}^{0}(\Omega)-E_{k}^{0}(\Omega)} \nonumber \\
   &=& \frac{1}{4 \hbar
  \Omega_n} ( \frac{b \hbar^2}{4m^2 \Omega_{n}^{2}} )^{2}
  \frac{64n^5 160n^4 -336n^3 -664n^2 -28n -24}{(2n + 1)^2} \ \ .
\end{eqnarray}
Here, we used Eq.(12) to express $\Omega_{n}^{2} - \omega_{n}^{2}$
in terms of $\Omega_n$ and $b$. Collecting terms, we obtain
\begin{eqnarray}
 E_n &=&  \frac{\hbar \Omega_n}{2}(2n + 1) - \frac{3b \hbar^2}{4m^2
          \Omega_{n}^2}(2n^2 + 2n +1)  \nonumber \\
      &+&\frac{1}{4\hbar \Omega_n}(\frac{b \hbar^2}{4m^2
       \Omega_{n}^2} )^{2} \frac{64 n^5 + 160 n^4 - 336 n^3 - 664 n^2
       - 28 n -24 }{(2n + 1 )^2} \ \ .
\end{eqnarray}
Now we can readily evaluate any energy eigenvalues of the anharmonic
oscillator given by Eq.(6).
First, we obtain $\Omega_n$ from Eq.(12) for a given $n$ and substitute
this value to Eq.(14).

In order to compare the present result with the conventional perturbation
and the variational calculations, we carry out numerical calculations
up to the second order using Eq.(14).
For this purpose, we choose $\frac{m \omega^2}{2} = 0.5 eV \AA^{-2}$
and carry out calculations for various values of $b$. The results
for the ground state energy are shown in Table I. The exact ground
state energy is obtained numerically using the Runge-Kutta-Fehlberg
algorithm \cite{nine}.
The result shows that the present method is clearly superior to the
conventional perturbation theory and the variational calculation.
First of all, the present method gives highly accurate values
in the regime where the conventional perturbation theory is not
applicable. This is because the perturbational basis has been
renormalized through the variational process to yield a better
convergence.
This renormalization effect can be also seen from the changing values
of $ \frac{1}{2} m \Omega^2$. For large $b$, $ \frac{1}{2} m \Omega^2$
becomes much larger than $ \frac{1}{2} m \omega^2$, thus making the
perturbation Hamiltonian, $H' = -\frac{m(\Omega^2 - \omega^{2}) }{2} x^2
+ b x^4$ correspondingly smaller.
Now, in order to examine this behavior in detail, we show, in Table II,
the results of the first and the second order calculations of the
approximations for $b = 0.05 eV \AA^{-4}$.
First, we observe that in the ordinary perturbation theory, convergence
is bad already. The first order result of the present approximation
scheme is same to the variational calculation as shown in Eq.(3) and (5).
We can see clearly that the renormalized perturbation calculation yields
a good convergence.

We now extend the calculation to the first excited state energy.
The results are shown in Table III. For $b = 0.05 eV \AA^{-4}$,
we obtain $E_1 = 5.092412 eV$ which is 100.02\% of the true value.
Here, the variational parameter $\frac{1}{2}m \Omega_{1}^2$ is
$0.990354 eV \AA^{-2}$ and clearly different from
$\frac{1}{2}m \Omega_{0}^2$. We note that the agreement for the first
excited state is again excellent. As mentioned earlier, the value of
the variational parameter, $\Omega$, is different for each level.

\section{Helium Atom}
We now apply the present method to evaluate the energy eigenvalues of
$He$ atom. Evaluation of the ground state energy using both the
standard perturbation method and the variational calculation are
given in any standard quantum mechanical
text books \cite{one,two,three,four}. It is known that the
standard perturbation calculation does not yield satisfactory result
because of the large coulomb interaction between electrons.
The variational approximation gives a better result, which will be
utilized below.

The Hamiltonian for a $He$ atom is given by
\begin{eqnarray}
  H = \frac{p_{1}^2}{2m} + \frac{p_{2}^2}{2m} - \frac{Ze^2}{r_1}
     - \frac{Ze^2}{r_2} + \frac{e^2}{r_{12}} \ \ ,
\end{eqnarray}
where $Z$ is 2. In order to evaluate the energy eigenvalues
variationally, we choose a parent Hamiltonian
\begin{eqnarray}
  H_{Z^*} = \frac{p_{1}^2}{2m} + \frac{p_{2}^2}{2m} -
            \frac{Z^{*}e^2}{r_1} - \frac{Z^{*}e^2}{r_2} \ \ ,
\end{eqnarray}
where $Z^*$ is the variational parameter \cite{one}.
With this choice, the perturbation Hamiltonian $H'$ is given by
\begin{eqnarray}
  H' = - \frac{(Z-Z^{*} ) e^2}{r_1}
     - \frac{(Z-Z^{*} ) e^2}{r_2} + \frac{e^2}{r_{12}} \ \ ,
\end{eqnarray}
and $H = H_{Z^*} + H'$. The ground state eigenfunction for this
parent Hamiltonian is given by
\begin{eqnarray}
  \Phi_{0}(Z^*) = \phi_{100}(Z^{*}, r_{1}) \phi_{100}(Z^{*}, r_{2})
                  \chi^{-} \ \ ,
\end{eqnarray}
where $\chi^{-} = \frac{1}{\sqrt{2}}
( |\uparrow \ \rangle |\downarrow \ \rangle
 - |\downarrow \ \rangle |\uparrow \ \rangle )$.
The expectation value of  $ \langle  \Phi_{0}(Z^*) | H |  \Phi_{0}(Z^*)
\rangle $ is readily expressed as a function of $Z^*$ \cite{one},
\begin{eqnarray}
  \langle  \Phi_{0}(Z^*) | H |  \Phi_{0}(Z^*) \rangle =
  - \frac{1}{2} m c^{2} \alpha^{2} (4Z^* Z - 2 {Z^*}^{2} - \frac{5}{4}
  Z^{*} ) \ \ .
\end{eqnarray}
The optimal value of $Z^*$ is obtained by minimizing Eq.(19) and given
by $Z^{*} = 1.6875$.
Substituting the above results to Eq.(3), we obtain
\begin{eqnarray}
  E_g = &-& \frac{1}{2} m c^{2} \alpha^{2} (4Z^* Z - 2 {Z^*}^{2}
        - \frac{5}{4} Z^{*} ) \nonumber \\
        &+& \sum_{k} \frac{| \langle  \Phi_{k}(Z^*) | H'
          |  \Phi_{0}(Z^*) \rangle |^2}{E_{g}^{0}(Z^{*})-
          E_{k}^{0}(Z^{*})} \nonumber \\
        &+& \cdot \cdot \cdot
\end{eqnarray}
The first term gives the usual variational result, $E_g = -5.6953 ryd$
for the ground state energy which is 98.077\% of the experimental value
of -5.8070 \cite{one}.
In calculating the second order contribution, we only have to consider
excited state wavefunctions which have antisymmetric spin parts,
since $ \Phi_{0}(Z^*) $ is antisymmetric in spins.
The general form of the wavefunction with antisymmetric spin parts
is given by
\begin{eqnarray}
  \Phi_{n' l' m'}^{n l m}(r_{1}, r_{2}) = A [ \phi_{nlm}(r_{1})
  \phi_{n'l'm'}(r_{2}) +
  \phi_{n'l'm'}(r_{1}) \phi_{nlm}(r_{2}) ] \chi^{-} \ \ ,
\end{eqnarray}
where $A = \frac{1}{\sqrt{2}}$ if $ n l m \neq n' l' m'$ and
$\frac{1}{2}$ if $n l m = n' l' m'$. For calculation of the second
order term, we divide the perturbation Hamiltonian, $H'$, into two
parts
\begin{eqnarray}
  H_{1}' = - \frac{(Z-Z^{*} ) e^2}{r_1}
           - \frac{(Z-Z^{*} ) e^2}{r_2} \ \ , \nonumber
\end{eqnarray}
and
\begin{eqnarray}
  H_{2}' = \frac{e^2}{r_{12}} \ \ .
\end{eqnarray}
The $H_{1}'$ contribution does not have any angular dependence.
Therefore, it can be straightforwardly calculated to yield
\begin{eqnarray}
  \langle \Phi_{n' l' m'}^{n l m} &|& H_{1}' \ \ | \ \Phi_{1 0 0}^{1 0 0}
  \ \  \rangle \nonumber \\
  &=& -2A (Z-Z^{*}) e^{2} \delta_{lm,00} \delta_{l'm',00}
     (X_{n} \delta_{n',1} + X_{n'} \delta_{n,1}) \ \ ,
\end{eqnarray}
where
\begin{eqnarray}
  X_{n} = \int dr r^{2} R_{n0}(r) \frac{1}{r} R_{10}(r) \ \ . \nonumber
\end{eqnarray}
The $H_{2}'$ contribution can be calculated by expanding
$ \frac{1}{|r_{1} - r_{2}|}$ in terms of spherical harmonics \cite{three}.
A lengthy but straightforward calculation yields the result
\begin{eqnarray}
   \langle \Phi_{n' l' m'}^{n l m} | \ \ H_{2}' \ \ | \ \Phi_{1 0 0}^{1 0 0}
  \ \ \rangle
  = \frac{2A e^{2} (-1)^{m}}{2l +1} \delta_{l,l'} \delta_{m,-m'}
      Y_{n n' l} \ \ ,
\end{eqnarray}
where
\begin{eqnarray}
  Y_{n n' l} = \int dr dr_{2} r_{1}^{2} r_{2}^{2} R_{nl}(r_{1})
   R_{n'l}(r_{2}) \frac{r_{<}^{l}}{r_{>}^{l+1}} R_{10}(r_{1})
   R_{10}(r_{2}) \ \ .
\end{eqnarray}
Collecting the above results, we obtain an expression for the ground
state energy,
\begin{eqnarray}
  E_g = &-& \frac{1}{2} m c^{2} \alpha^{2} (4Z^* Z - 2 {Z^*}^{2}
        - \frac{5}{4} Z^{*} ) \nonumber \\
        &+& \sum_{n,n',l,m} \frac{| \langle  \Phi_{n' l -m}^{nlm} | H'
          |  \Phi_{100}^{100} \rangle |^2}{-\frac{1}{2}m c^2
          (Z^{*} \alpha )^{2} [ 2 - \frac{1}{n^{2}} -
           \frac{1}{{n'}^{2}}]} \ \ ,
\end{eqnarray}
where
\begin{eqnarray}
  | \langle  \Phi_{n' l -m}^{nlm} &|& H' \
          | \ \Phi_{100}^{100} \ \ \rangle |^2  \nonumber \\
    &=& | -2A ( Z - Z^* ) e^2 \delta_{lm,00} ( X_n \delta_{n',1}
     + X_{n'} \delta_{n,1} ) \nonumber \\
     &+& \frac{2A e^{2} (-1)^{m}}{2l +1} \delta_{lm,l'-m'}
      Y_{n n' l} |^2 \ \ .
\end{eqnarray}
Here, we assume $n' \geq n$, $0 \leq l \leq n-1$, and $0 \leq m \leq l$.
Also $n'=n=1$ should not be included in the second order correction.
The above expression is numerically evaluated upto $n'=7$.
The second order correction gives $-0.0249 ryd$, and, thus, makes
the total ground state energy $-5.7202 ryd$, which is 98.505\% of the
experimental value of -5.8070.
Comparing to the original variational calculation value of -5.6953,
we observe that a systematic improvement has been achieved, although
the agreement is not as good as the case of anharmonic oscillator.
We believe that this rather slow convergence may originate from the fact that
we used the same $Z^*$ for both electrons. It is known that if one electron
draws closer to the nucleus at some instant, it tends to push the other
farther out, thus making the screened charges different for the two
electrons \cite{one}. Also using the same $Z^*$ for excited states may
introduce additional errors.
However, either introducing an additional variational parameter or
calculating $Z^*$ for each excited level would make the
present scheme too complicated to be useful.
We just note that for the first excited state of symmetric spin state, $Z^*$
is given by 1.8497 which is quite different from that of the ground
state. The first excited state energy obtained variationally without
the second order correction is given -4.2765 $ryd$ comparing to the
experimental value of -4.3504 $ryd$. Because of computational
complexity, we have not carried out the full second order calculation.
However, partial calculations with some selected terms also show the
same trends as in the case of the ground state.

\section{Conclusions}
We have presented an approximation scheme which combines the perturbation
theory with the variational calculation in quantum mechanics.
It is shown that this scheme provides a very good convergence beyond the
first variational calculation even when the perturbing potential is large.
Also it can be readily used to estimate excited states. Therefore, this
method overcomes the shortcomings of the perturbation theory and the
variational calculation and combines the advantages, when applicable.
Here, it should be noted that the present method is applicable only when
a parent Hamiltonian can be found, thus limiting its usefulness.
However, it is known that for a quite large number of problems, potentials
can be expanded into simple harmonic potential plus higher order terms.
In such category of problems, the present method is expected to be efficient.

Another merit of the present method which, we believe, may be more important
is the pedagogical value of the combined approximation scheme. The theory
shows to the students the limits of both approximations and demonstrates a
way to overcome those, although in a limited class of problems.
Also, this method provides an easy example of renormalized perturbation
theory which are often used in many body and field theories \cite{ten}.

\acknowledgements

This work has been partially supported by the Korean Ministry of
Education (BSRI-97-2425),
the Korea Science and
Engineering Foundation through Project No. 95-0701-04-01-3 and also
through the SRC program of SNU-CTP, and Yonsei University
Research Fund.

\newpage
\begin{table}
\caption{Comparison of the ground state energies(in eV) obtained using
         different approximation schemes.The values in the parentheses
         are the ratios to the exact values.
         The values of $\frac{1}{2} m \Omega_{0}^2$ are also
         shown as references.}
\item[]\begin{tabular}{@{}cc|cc|cc|cc}
 & b(eV$\AA^{-4}$)     & 0.01     & & 0.05     & & 0.25   &  \\ 
 \tableline
 & Perturbation        & 1.4318427 & & 1.5279252 & & does not converge.  & \vspace{-0.2cm} \\
  & Theory              & (99.935\%) & & (95.962\%) & &   & \\ 
 \tableline
 & Variational         & 1.4333279 & & 1.5968858  & & 2.0664772 &  \vspace{-0.2cm} \\
 & Calculation         & (100.038\%) & & (100.293\%) & & (100.929\%) & \\ 
 \tableline
 & Present             & 1.4327276   & & 1.5912088  & & 2.0412648  &  \vspace{-0.2cm} \\
 & Method              & (99.997\%)  & & (99.937\%)  & & (99.679\%)  & \\  
 \tableline
 & Exact               & 1.4327725  & & 1.5922195  & & 2.0474629  &   \vspace{-0.2cm} \\
 & Energy Value        &           & &           & &         &  \\ 
 \tableline \tableline
 & $\frac{1}{2} m \Omega_{0}^2 (eV \AA^{-2})$ & 0.5770839 & & 0.8227827 & & 1.6423320 & \\
\end{tabular}
\end{table}

\begin{table}
\caption{Detailed comparison of the ground state energy values (in eV)
         obtained by different methods for b=0.05 eV$\AA^{-4}$.}
\item[]\begin{tabular}{@{}clc|cc|cc}
&Method of     &  & First order & & Second order & \vspace{-0.2cm}  \\
&Approximation &  & calculation & & calculation  &  \\ 
\tableline \tableline
&Perturbation  &  & 1.6659633   & & 1.5279252    & \vspace{-0.2cm}  \\
&              &  & (104.632\%) & & (95.962\%)   & \\ 
\tableline
&Variational   &  & 1.5968858   & & ------           & \vspace{-0.2cm}  \\
&Calculation   &  &(100.293\%)  & &            & \\ 
\tableline
&Present       &  &1.5968858    & & 1.5912088    &  \vspace{-0.2cm} \\
&Method        &  & (100.293\%) & & (99.937\%)   & \\
\end{tabular}
\end{table}

\begin{table}
\caption{Comparison of the first excited state energies (in eV) for
         b=0.05eV$\AA^{-4}$. The values in the parentheses are
         ratios to the exact value. The value of $\frac{1}{2} m \Omega_{1}^2$
          is shown together as a reference. }
\item[]\begin{tabular}{@{}clc|cc}
&Perturbation  & & 4.484801(88.09\%) & \\ 
\tableline
&Variational Calculation & & 5.106102(100.29\%) & \\ 
\tableline
&Present Method  & & 5.092412(100.02\%) & \\ 
\tableline
&Exact Value  & & 5.091282 & \\ 
\tableline \tableline
&$\frac{1}{2} m \Omega_{1}^2 (eV \AA^{-2})$ & & 0.990354 & \\
\end{tabular}
\end{table}

\end{document}